\documentclass[twocolumn,showpacs,preprintnumbers,amsmath,amssymb]{revtex4}
\usepackage{graphicx}
 
\usepackage{dcolumn}
 
\usepackage{bm}

\newcommand{\lsco}{La$_{2-x}$Sr$_{x}$CuO$_4$}

\newcommand{\srruo}{Sr$_2$RuO$_4$}
\newcommand{\srdzs}{Sr$_3$Ru$_2$O$_7$}

\newcommand{\casrruo}{Ca$_{2-x}$Sr$_{x}$RuO$_4$}

\newcommand{\caruoef}{Ca$_{1.5}$Sr$_{0.5}$RuO$_4$}
\newcommand{\srrutio}{Sr$_2$Ru$_{1-x}$Ti$_x$O$_4$}
\newcommand{\gbd}{$\gamma$-band}
\newcommand{\abd}{$\alpha$-band}
\newcommand{\bbd}{$\beta$-band}
\newcommand{\gsh}{$\gamma$-sheet}
\newcommand{\ash}{$\alpha$-sheet}
\newcommand{\bsh}{$\beta$-sheet}

\newcommand {\Qi}{$\mathbf{q_i}$}

\newcommand {\om}{$\omega$}

\newlength{\figwidth}
 
\divide\figwidth by 2
 
\begin{document}

\title{Inelastic neutron scattering study of magnetic excitations
in Sr$_2$RuO$_4$.}
\author{M.~ Braden$^{a,b,c,*}$,Y. ~Sidis$^b$, , P. ~Bourges$^b$,
P. Pfeuty$^b$, J. Kulda$^d$, Z. Mao$^e$, Y. ~Maeno$^{e,f}$}  
\address{
$^a$ Forschungszentrum Karlsruhe, IFP, Postfach 3640, D-76021 
Karlsruhe, Germany\\
$^b$Laboratoire L\'eon Brillouin,
C.E.A./C.N.R.S., F-91191-Gif-sur-Yvette CEDEX, 
France\\
$^c$ II. Phys. Inst., Univ. zu K\"oln, Z\"ulpicher Str. 77, D-50937 K\"oln, Germany\\
$^d$ Institut Laue-Langevin, Boite Postale 156, 38042 Grenoble Cedex 9, 
France\\
$^e$ Department of Physics, Kyoto University,
Kyoto 606-8502, Japan\\
$^f$ International Innovation Center (IIC), Kyoto 606-8501, Japan
}
\date{\today}


\begin{abstract}
Magnetic excitations in \srruo ~ have been studied 
by inelastic neutron scattering. The magnetic 
fluctuations are  dominated by incommensurate peaks related to the 
Fermi surface nesting of the quasi-one-dimensional 
$d_{xz}$- and $d_{yz}$-bands. 
The shape of the incommensurate signal agrees well
with RPA calculations.
At the incommensurate {\bf Q}-positions
the energy spectrum considerably softens upon cooling
pointing to a close magnetic instability : 
\srruo ~does not exhibit quantum criticality
but is very close to it. $\omega / T$-scaling may be fitted
to the data for temperatures above 30\ K.
Below the superconducting transition, the magnetic response
at the nesting signal is not found to change in the 
energy range down to 0.4meV.

\end{abstract}


\maketitle

\section{INTRODUCTION}

\srruo ~ is still the only superconducting layered perovskite besides the
cuprates\cite{maeno}; however, in contrast to the cuprate high temperature
superconductors (HTSC), superconductivity in \srruo ~ develops in a
well defined Fermi-liquid state \cite{maeno,2}.
Nevertheless the superconducting state and the pairing
mechanism in \srruo ~ are highly unconventional.
The present interest in this compound goes far beyond the
simple comparison with the cuprate high temperature superconductors.

The extreme sensitivity of the superconducting transition
temperature on non-magnetic impurities suggests a non-phonon
mechanism \cite{3}. It is  further established that
superconductivity in \srruo ~ is made of spin-triplet Cooper pairs
and breaks time-reversal symmetry \cite{2,4}. 
The strongest experimental argument for that can be found
in the uniform susceptibility
measured either by the NMR-Knight-shift or polarized neutrons experiments
 \cite{5,hayden}
and in the appearance of spontaneous fields 
in the superconducting state reported by $\mu$SR \cite{6}.
A spin-triplet 
p-wave order parameter had been proposed before these experiments \cite{7},
in the idea that superconductivity arrises from
a dominant interaction with ferromagnetic fluctuations
in analogy to superfluid $^3$He.
Rice and Sigrist stressed the comparison with the
perovskites SrRuO$_3$ and CaRuO$_3$ which order ferromagnetically 
or are nearly ferromagnetic respectively \cite{srruo3}.
Evidence for ferromagnetic fluctuations in \srruo ~ was inferred
from NMR-experiments : Imai et al. found that $1 \over {T_1T}$ of
the O- and of the Ru-NMR exhibit a similar temperature
dependence and interpreted that this could be only due to
ferromagnetic fluctuations \cite{imai}.

The macroscopic susceptibility in \srruo ~ is enhanced
when compared with the band structure calculation but only weakly,
in particular its temperature dependence remains flat \cite{2,9}.
There exist also layered ruthenates which are very close
to ferromagnetic order at low temperatures : the purest \srdzs ~ samples
show meta-magnetism and samples with somehow less quality even order
ferromagnetically \cite{10}.
A highly enhanced susceptibility pointing to a ferromagnetic instability
is also observed in the phase diagram of \casrruo ~ \cite{11,12} 
but for a rather high Ca concentration,
\caruoef ~\cite{13}.
In these nearly ordered layered ruthenates,
the susceptibility is about two orders of magnitude
higher than that in \srruo ~ and strongly
temperature dependent.

Some doubt about the unique role of ferromagnetism in \srruo ~
arose from the strong moment antiferromagnetic order observed in
the Ca analog \cite{14}, 
which inspired Mazin and Singh to perform
a calculation of the generalized susceptibility
based on the electronic band structure \cite{15}.
Surprisingly they found that the dominating part is neither
ferro- nor antiferromagnetic but incommensurate.
The Fermi-surface in \srruo ~ is well studied both by experiment \cite{16}
and by theory \cite{17} with satisfactory agreement.
Three bands are contributing to the Fermi-surface which may be
roughly attributed to the three t$_{2g}$-levels,
the d$_{xy}$-, d$_{yz}$- and d$_{xz}$-orbitals, 
occupied by the four
4d-electrons of Ru$^{4+}$.
The d$_{xy}$-orbitals hybridize in the xy-plane and,
therefore, form a band with two-dimensional character, \gbd .
In contrast,
the d$_{xz}$ and the d$_{yz}$ orbitals may hybridize only along
the x and the y-directions, respectively, and form quasi-one-dimensional
bands, \abd ~ and \bbd, with flat sheets in the 
Fermi-surface, \ash ~ and \bsh . 
The latter give rise to strong nesting
and enhanced susceptibility for ${\bf q}=(0.33,q_y,0)$  or 
${\bf q}=(q_x ,0.33,0)$ \cite{15}. Along the
diagonal both effects strengthen each other yielding a pronounced
peak in the susceptibility at ${\bf q_{i-cal}}=(0.33,0.33,0)$.
Using inelastic neutron scattering (INS)  we have perfectly confirmed this
nesting scenario \cite{sidis}. 
The dynamic susceptibility at moderate energies
is indeed dominated by the incommensurate fluctuations occurring
very close to the calculated position at ${\bf q_i}=(0.30,0.30,0)$.
Furthermore, we found a pronounced temperature dependence which
can explain the temperature dependence of both Ru- and 
O-$1 \over T_1T$-NMR \cite{imai}.

In the meanwhile, several groups have worked on theories where
superconductivity is related to the
incommensurate fluctuations \cite{18,19,20}. These theories,
however,
do not yet give an explanation for the fine structure of the
order parameter. Specific heat data on the highest quality samples
clearly indicates the presence of line nodes in the superconducting
state \cite{21}. Ultrasonic \cite{22}
and thermal conductivity \cite{23} results would disagree
with vertical line nodes which leave horizontal
line nodes as the only possible ones. Zhitomirsky
and Rice \cite{24} assume that superconductivity is transferred from
the \gbd ~ to the \abd ~ and \bbd ~ by a proximity effect
and get a conclusive explanation for the horizontal line nodes.
In this model superconductivity should be mainly related to
excitations associated with the active \gbd , which so far
have not been characterized.
Therefore, it appears still interesting to further
deepen the study of the magnetic fluctuations in \srruo ~.

In this paper we report on additional INS
studies in \srruo ~ in the normal as well as in the superconducting
phase. We present a more quantitative analysis of the
incommensurate fluctuations related to the \ash  ~ and \bsh ~
and discuss the possible contributions due to the two-dimensional
\gbd .

\section{Experimental}

\subsection{Experimental setup}

Single crystals of \srruo ~ were grown by a floating zone method in an
image furnace; they exhibit the superconducting transition at
T$_c$=0.7, 1.35 and 1.43K and have volumes of about 450mm$^3$ each.
Since most of the measurements were performed
in the normal phase, where the differences in T$_c$ should not
affect the magnetic excitation spectrum,
we aligned the three crystals together in order
to gain counting statistics in the INS
experiments.
Count rates in the ruthenates experiments are relatively small
already due to the steeper decrease of the Ru magnetic
form-factor with increasing scattering vector.
For the measurements below and across the superconducting transition
we mounted only the two crystals with relatively high T$_c$ together.
The mounting of the two or three crystals was achieved with individual
goniometers yielding a total mosaic spread below 0.5 degrees.

We used the thermal triple axis spectrometer 1T installed at the
Orph\'ee-reactor (Saclay, France)
in order to further characterize the scattering in the
normal state. The instrument was operated with double focusing
pyrolithic graphite (PG) monochromator and analyzer, in addition
PG filters in front of the analyzer were used to suppress higher order
contamination, the final neutron energy was fixed at 14.7meV.
All diaphragms determining the beam paths were opened more widely
than usually in order to relax ${\bf Q}$-resolution,
since the magnetic signals are not sharp in ${\bf Q}$-space.
In most experiments, the scattering plane was defined by (1,0,0) and (0,1,0) directions
in order to span any directions within the  (ab) plane. An additional experiment
has been made with the (1,1,0) and (0,0,1) directions with the plane to 
follow the spin fluctuations along the c$^*$ axis.

Studies aiming at the response in the magnetic
excitation spectrum on the opening of the superconducting gap
require an energy resolution better than the expected value for
twice the superconducting gap. Therefore, such experiment is better placed
on a cold triple-axis spectrometer even though this implies a sensitive
reduction in the flux. We have made preliminary studies using the
cold spectrometers 4F at the Orph\'ee reactor and a more
extensive investigation on the spectrometer IN14 at the ILL.
These instruments possess PG monochromators
(double at 4F and focusing at IN14)
and focusing analyzers. 
The final neutron energy was 
fixed at $E_f$= 5 meV on both cold source spectrometers where 
a Be filter has been employed to cut down higher-order wavelength neutrons.
Cooling was achieved by use of a dilution-
and a He$^3$-insert at 4F and IN14 respectively.

\subsection{Theoretical background for magnetic neutron scattering}

The magnetic neutron cross section per 
formula unit is written in terms of
the Fourier transform of the spin correlation function, $\rm S_{\alpha\beta}({\bf Q}, \omega)$
(labels $\alpha,\beta$ correspond to x,y,z) as \cite{Lovesey};

\begin{equation}
\frac{d^2 \sigma}{d \Omega d \omega}= \frac{k_f}{k_i} r_0^2 
F^2({\bf Q})
\sum_{\alpha,\beta} (\delta_{\alpha,\beta}-\frac{Q_{\alpha}Q_{\beta}}{|{\bf Q}|^2})
S_{\alpha\beta}({\bf Q}, \omega)
\; \label{eq:INS}
\end{equation}
where k$_i$ and k$_f$ are the incident and final neutron wave vectors, 
$r_0^2$=0.292 barn, $F({\bf Q})$ is the magnetic form factor.
The scattering-vector ${\bf Q}$ can be split into 
${\bf Q}$ =${\bf q}$+${\bf G}$, where ${\bf q}$ lies in the 
first Brillouin-zone 
and ${\bf G}$ is a zone-center. 
All reciprocal space coordinates $(Q_x,Q_y,Q_z)$ are given in reduced lattice
units of 2$\pi$/a or 2$\pi$/c.

According to the fluctuation-dissipation theorem, 
the spin correlation function is related to the imaginary part of the dynamical
magnetic susceptibility times the by one enhanced Bose-factor:
\begin{equation}
 S_{\alpha\beta}({\bf Q}, \omega)=\frac{1}{\pi (g  \mu_B)^2} \frac{\chi_{\alpha\beta}"({\bf Q},\omega)}
{1-\exp(-\hbar \omega / k_BT)}
\; \label{eq:INSchi}
\end{equation}
In case of weak anisotropy, which is usually observed in a
paramagnetic state, $\rm \chi_{\alpha\beta}"({\bf Q},\omega)$ reduces to 
$\rm \chi"({\bf Q},\omega) \delta_{\alpha,\beta}$. 
Note that for itinerant magnets,
anisotropy of the susceptibility can occur due to spin-orbit coupling.
$\rm F({\bf Q})$ can be described by the Ru$^+$ magnetic form factor
in first approximation. 
Once determined, the magnetic response is converted to the dynamical susceptibility
and calibrated in absolute units by comparison with acoustic phonons,
using a standard procedure depicted in \cite{sidis}. 
 
\section{Results and discussion}

\subsection{ RPA analysis of the magnetic excitations }  

At low temperature
\srruo ~ exhibits well defined Fermi-liquid properties; therefore,
it seems appropriate to assess its magnetic excitations within
a RPA approach basing on the band structure.
Density functional calculations in LDA were performed by several
groups and yield good agreement with the Fermi-surface determined
in de-Haas van Alphen experiments. 
The bare non-interacting susceptibility, $\chi ^0({\bf q})$,
can be obtained by summing the matrix elements for an electron
hole excitation  \cite{Lovesey}:
\begin{equation}
\chi^0 ({\bf q}, \omega) = -  \sum_{{\bf k},i,j}
M_{{\bf k}i,{\bf k+q}j}
\frac{f(\varepsilon_{{\bf k+q},j})-f(\varepsilon_{{\bf k},i})}
{\varepsilon_{ {\bf k+q},j}-\varepsilon_{{\bf k},i}-\hbar \omega + i \epsilon}
\; \label{eq:1}
\end{equation}
where $\epsilon \rightarrow$0, $f$ is the Fermi distribution function 
and $\varepsilon_k$ 
the quasiparticle dispersion relation. 
This was first calculated by Mazin and Singh \cite{15}
under the assumption that only excitations within the same 
orbital character are relevant (the matrix-elements $M_{{\bf k}i,{\bf k+q}j}$
are equal to one or zero).
Mazin and Singh predicted the existence of peaks in the real part of the
bare susceptibility at \om =0 due to the pronounced nesting
of the \abd s and \bbd s. 
These peaks were calculated at (0.33,0.33,0)
and experimentally confirmed very close to this position at \Qi =(0.3,0.3,0),
see Fig. 1.
In addition to the peaks at \Qi , this study
finds ridges of high susceptibility at (0.3, $q_y$,0) for
$0.3<q_y<0.5$ and some shoulder for $0<q_y<0.3$.

The susceptibility gets enhanced through the Stoner-like
interaction which is treated in RPA by :
\begin{equation}
\chi ({\bf  q} ) = { \chi ^0 ({\bf q}) \over 1-I({\bf q})\chi^0({\bf q})} 
\; \label{eq:2}
\end{equation}
with the ${\bf q}$-dependent interaction $I({\bf q})$.
For the nesting positions Mazin and Singh get $I({\bf q})\chi^0({\bf q})$=1.02,
which already implies a diverging susceptibility and
a magnetic instability.

In Fig. 1 we show a scheme of the (hk0)-plane in reciprocal
space. Due to the  body centering in space-group I4/mmm any
(hkl)-Bragg-points have to fulfill the condition (h+k+l) even;
(100) for instance is not a zone-center but a Z-point. The boundaries of the
Brillouin-zones are included in the figure. Large filled circles
indicate the position of the incommensurate peaks as predicted
by the band-structure and as observed in our previous work. The
thick lines connecting four of them correspond to the ridges of
enhanced susceptibility also suggested in reference \cite{15}.

In the meanwhile several groups have performed similar calculations
which all agree on the dominant incommensurate fluctuations
\cite{20,24b,25,26,27}.
However there are serious discrepancies concerning
the detailed structure of the spin susceptibility away from \Qi.
These discrepancies mainly rely on the parameters used to describe the
electronic band structure, on the choice of
$\rm I({\bf q})$ and on the inclusion of more subtle
effects such as spin-orbit or Hund couplings.

\begin{figure}[tp]
\resizebox{0.8\figwidth}{!}{
\includegraphics*{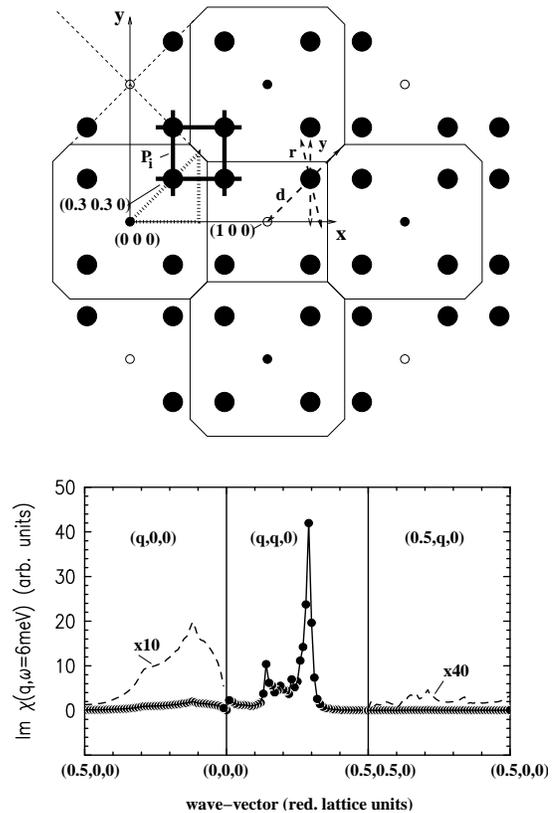}}
\caption{Upper part :
Scheme of the (hk0)-plane in reciprocal space.
Thin lines show the boundaries of the
Brillouin-zones and small filled and open circles
the zone-centers and the Z-points of the body centered lattice.
Large filled circles
indicate the position of the incommensurate peaks and
thick lines connecting four of them correspond to the walls of
enhanced susceptibility also suggested in reference [17].
The dashed double arrows 
show the paths of the constant energy scans frequently
used in this work: along [010]-direction -- {\bf y}-scan, transverse to the
${\bf Q}$-vector, {\bf r}-scan, and along the diagonals in [110]-direction,
{\bf d}-scans.
Lower part : Imaginary part of the generalized susceptibility calculated
by RPA along the three paths indicated in the upper part 
as dotted lines.}
\label{fig1}
\end{figure}

%

In order to compare directly to the INS experiments, see 
equations (1) and (2), it is necessary to perform the RPA analysis 
by taking into
account both real and imaginary part of the susceptibility.
Morr et al. \cite{27} report such calculations obtained 
by fitting the band structure to the ARPES results \cite{28}.
They find in addition to the peak at \Qi ~ quite strong intensity
near ${\bf P_i}$=(0.3,0.5,0), (the middle of the walls, see Fig. 1), which   is
even comparable to that at \Qi ~ in the bare susceptibility.
Similar results were obtained in references \cite{20,24b}.
Ng and Sigrist \cite{25} find much less spectral
weight in the ridges at (0.3, $q_y$) for
$0.3<q_y<0.5$ but stronger shoulders $0<q_y<0.3$.
In addition, they calculate the separate contribution
of the \gbd ~ which does not show a particular enhancement
in the ferromagnetic ${\bf q}$-range but is little structured.
Eremin et al. \cite{26} calculate the susceptibility taking into account 
strong
hybridization and obtain results somehow differnt from the
other groups.
They find a strong signal at ${\bf P_i}$;
in addition there is some enhancement of the susceptibility
related to the van-Hove singularity of the \gbd .
This contribution occurs quite close
to the zone center at ${\bf q_{vH}}$=(0.15,0,0).

We have performed the full RPA analysis basing on the LDA band
structure reported in reference \cite{17,15}  
in order to accompany our
experimental investigations.
We first calculate the bare electron hole susceptibility $\chi^0$ 
from the usual expression, equation (3). 
For the band energies $\epsilon(k_x,k_y,k_z)$
we use the expressions of Mazin and Singh \cite{17} for the three
 mutually non hybridizing tight-binding bands in the vicinity of the 
Fermi level:
\begin{eqnarray}
\epsilon_{xy}(k_x,k_y,k_z) & = & 400mev(-1+2(cos(k_x)+cos(k_y))\nonumber \\
                           &   &   -1.2cos(k_x)cos(k_y))\\
\epsilon_{xz}(k_x,k_y,k_z) & = & 400mev(-.75+1.25cos(k_x)-\nonumber\\
                           &   & .5cos(k_x/2)cos(k_y/2)cos(k_z/2))\\
\epsilon_{yz}(k_x,k_y,k_z) & = & 400mev(-.75+1.25cos(k_y)-\nonumber\\
                           &   & .5cos(k_x/2)cos(k_y/2)cos(k_z/2)) 
\end{eqnarray}
We also used the crude approximation that matrix-elements for transitions 
between bands of the same character are equal to one and others zero.
The {\bf q}-dependent "ferromagnetic" interaction (Stoner factor) I({\bf q}) is taken
to be equal to (following reference \cite{15}): 
$I({\bf q})=\frac{320mev}{1+0.08 {\bf q}^2}$.
With this choice of I({\bf q}) the calculated static susceptibility 
$\chi({\bf q}=0,\omega=0)$
is slightly lower than the measured one. 
If I({\bf q}) is chosen larger an instability appears
at the incommensurate wave vector.

Our results for the imaginary part of the generalized
susceptibility at an energy transfer of 6meV are given 
in the lower part of figure 1.
Besides the dominating nesting peak near \Qi ~ there is 
a further contribution near (0.15,0.15,0) which is related
to the \gsh . In contrast we find a small susceptibility near 
${\bf P_i}$ and for ($q_x$,0,0) with small values of $q_x$.

Since the 
${\bf q}$-position of the magnetic excitations were found not to depend
on energy, it is easiest to observe the signal in INS by scanning at constant
energy. The scan paths are included in Fig. 1, they are purely transverse
or rocking-like, {\bf r}-scan, along a [100]-direction, {\bf x}-scans,
and in diagonal direction, {\bf d}-scans. 

\begin{figure}[tp]
\resizebox{0.7\figwidth}{!}{
\includegraphics*{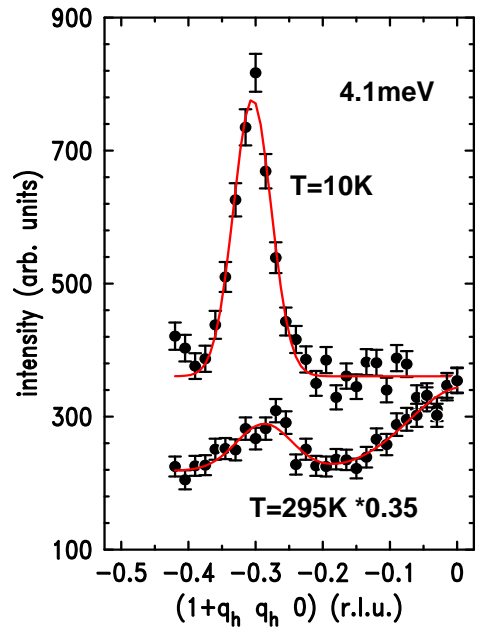}}
\caption{ {\bf d}-scans across the incommensurate peak $Q=(0.7,-0.3,0)$ at low and
high temperatures; the data at 295 K were scaled by 0.35.
The disappearance of the signal upon heating gets partially compensated by the gain 
through the Bose-factor.}
\label{fig2}
\end{figure}

The observed signal is rather broad and, therefore, the scans performed
are extremely wide covering complete cuts through the Brillouin-zone.
This further implies that the background (BG) may be non-constant at least
sloping. Also, the signals are relatively weak compared to typical
triple axis spectrometry 
problems; this implies that sample independent BG-contributions
which usually are negligible play a role. 
Fig. 2 presents the results of {\bf d}-scans at different
temperatures clearly demonstrating the gain in statistics compared to
the previous work \cite{sidis}.
The magnetic intensity shown in Fig. 2 disappears upon heating but
this effect gets partially compensated by the gain through the
Bose-factor.

\subsection{ Shape of the incommensurate signal}

The fact that the  incommensurate signal around a zone-center and
around a Z-point, (001), are equivalent already indicates that the coupling
between RuO$_2$-planes is negligible, i.e. that in-phase and out-of-phase
coupling are indistinguishable. The 2D-character has been directly documented
by Servant et al. \cite{29} who found no q$_l$-dependence at (0.3,0.3,$q_l$) 
for q$_l$ between -0.5 and 0.5. 
We find the same result by varying Q$_l$ in a broader
range between 2 and 5 in (0.3,0.3,Q$_l$), see Fig. 3.
This 2D character is actually surprising
since the dispersion relation of the quasi-particles involved in the
computation of $\rm \chi"({\bf q}, \omega)$ is not purely 2D \cite{15}.
One may therefore expect weak spin correlation along the $c^*$-direction.

Recently, it has been shown that these fluctuations freeze out
into a spin density wave (SDW) ordering by a minor replacement of Ru by Ti
\cite{srruti}.
In this ordering a very short correlation length along the c-direction
nicely reflects the 2D-character of the incommensurate inelastic signal.
The SDW propagation vector finally favored in the static ordering
corresponds to the out-of-phase coupling between neighboring layers.
This static interlayer-coupling might be even
due to the CDW always coupled to a SDW,
which, however, has not yet been discovered so far.
One may add that the stripe ordering in
La$_{1.67}$Sr$_{0.33}$NiO$_4$, 
which differs from the \srrutio-case in many aspects but
nevertheless may be considered as a mixed SDW-CDW-ordering,
occurs at the same propagation vector \cite{lsno}.

\begin{figure}[tp]
\resizebox{0.7\figwidth}{!}{
\includegraphics*{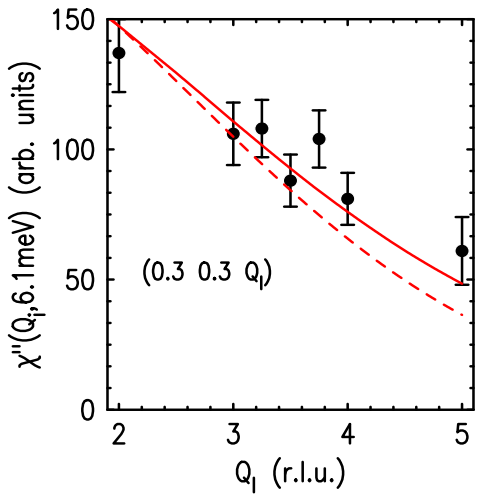}}
\caption{
Height of the magnetic signal at the incommensurate position
as function of $Q_l$; the dashed line shows the dependence 
expected from the Ru$^{1+}$-form factor and the solid line 
that assuming the spin-density distribution observed recently
in \caruoef \cite{34} , see text.}
\label{fig3}
\end{figure}

The wider Q$_l$-dependence of the incommensurate signal
shown in Fig. 3
can give information about the anisotropy of the
excitation, since INS measures only the spin component perpendicular to Q.
We consider the diagonal susceptibility 
$\chi"_{\alpha \beta}$ with tetragonal symmetry
$\chi_{\pm}:=\chi"_{xx}=\chi"_{yy}\not= \chi"_{zz}$.
The measured intensity is then given by :
\begin{equation}
\frac{d^2 \sigma}{d \Omega d \omega} \propto  
F^2({\bf Q})
\lbrack(1-\frac{Q_l^2}{|{\bf Q}|^2}) \chi"_{zz}
+
(1+\frac{Q_l^2}{|{\bf Q}|^2}) \chi"_{\pm}
\rbrack
\; \label{eq:INSanisotropy}
\end{equation}
which implies that the observations at high Q$_l$ favor the
in-plane component of the susceptibility. 
For the detailed analysis, one has to compare
with the form factor. In the figure we show the Q$_l$-dependence
assuming that spin density is localized at the Ru-site and may be
modeled by the form factor of Ru$^{1+}$ \cite{formfac}. 
The Ru-form-factor dependence underestimates
the signal at higher Q$_l$-values; however, the  Ru$^{1+}$-form factor 
is certainly a too crude approximation. Measurements of the spin-density
distribution induced by an external field have not been very precise
due to the small magnetic susceptibility and the resulting small moment
in \srruo ~\cite{32}.
However, in \caruoef ~  which exhibits ferromagnetic ordering below 1K 
and whose
low temperature susceptibility is about two orders of magnitude higher than
that in \srruo ~\cite{11,12,13}, it has been possible to study
the field induced spin-density distribution.
These experiments revealed 
an extremely high amount of spin-density at the oxygen position,
about one third of the total moment \cite{34} and an orbital 
contribution at the Ru-site.
By use of the \caruoef ~spin-density distribution one 
obtains a good description of the $Q_l$-dependence given in 
Fig. 3.
However, the form-factor in \srruo ~ should be even more complex.
Since the main contribution originates from 
the flat $d_{xy}$-orbitals \cite{26}, there must be an anisotropy in the
effective form factor which  indeed  
was observed in \caruoef \cite{34} . Qualitatively, the anisotropy in the
form-factor has to be compensated by some weak anisotropy in the
spin susceptibility, i.e. by an enhanced out-of-plane component.
Such susceptibility anisotropy corresponds to the orientation of the
spins in the SDW ordering phase in \srrutio ~ \cite{srruti}
where the spins are aligned parallel to the c-direction
and also to the conclusion deduced by Ng and Sigrist from the 
influence of spin-orbit coupling in \srruo \cite{25}.

\begin{figure}[tp]
\resizebox{0.7\figwidth}{!}{
\includegraphics*{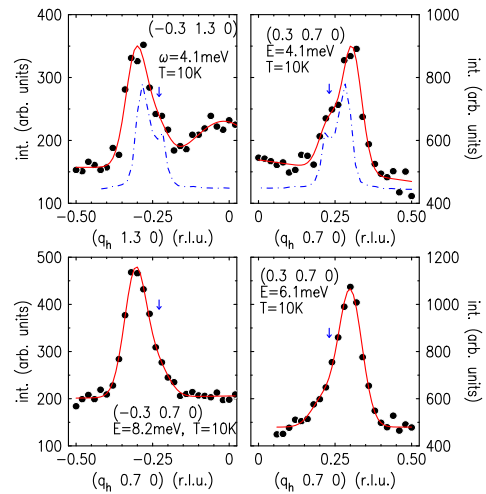}}
\caption{ Results from constant energy scans across the 
incommensurate position along the [100]-direction.
Solid lines denote fits with Gaussians and dashed lines
the calculated 
imaginary part of the generalized susceptibility at the energy of
4.1meV (shifted in y-axis). Arrows indicate the position of the shoulder
seen in the RPA calculations.}
\label{fig4}
\end{figure}

{\it -- Shoulder of the incommensurate peak --}
In order to examine
the shape of the incommensurate peak and to verify the existence
of the ridges of additional nesting intensity or the additional
peak ${\bf P_i}$, reported in different band structure analyzes
\cite{15,24b,25,26,27}, we have
scanned across \Qi ~ along the [100] or [010]-directions, {\bf y}-scans
see Fig. 1.
The results are shown in Fig. 4. One can recognize that the
incommensurate peak is not symmetric but exhibits always a shoulder
to the lower $q_x$-side in absolute units.
The shoulder is seen in many scans in reversed
focusing conditions of the spectrometer configuration
excluding
an experimental artefact.
Our full RPA calculation nicely agrees with such shoulder; the thin line
in Fig. 4 shows the calculated imaginary part of the susceptibility
at the energy transfer of the scan
and describes the observed signal perfectly besides a minor
offset in the position of the incommensurate signal. 
In contrast, neither the experiment
nor our full RPA analysis yield significant intensity in the ridges,
i.e. the range ($q_x$,0.3,0) with $0.3<q_x <0.5$, see Fig. 1 and 4.
In particular an intensity at $P_i$ only three times weaker than
that at \Qi ~ would have been easily detected experimentally.
The nesting peak appears to be isolated with two shoulders along
[100] and [010] to the lower absolute $q_{x,y}$-sides.
These shoulders are connected to the \gsh .

\subsection{ Possible additional magnetic excitations }

In none of the band-structure calculations there is evidence
for a strong and sharp enhancement of any susceptibility
exactly at the zone-center pointing to a ferromagnetic instability
\cite{24b,25,26,27,15}.
However, several of these calculations find some large susceptibility
near the zone-center which can be associated with the
van Hove-singularity of the \gbd ~ closely above the
Fermi-level. Eremin et al. \cite{26} report this signal at 
${\bf q}$=(0.1-0.2,0,0),
other groups find a small peak along the diagonal
(q,q,0) \cite{20,25,24b},. Our own analysis also yields such a signal
which is found to be strongest at (0.15,0.15,0) and which
levels out along the [100] and [010] directions,
see Fig. 1.

\begin{figure}[tp]
\resizebox{1.0\figwidth}{!}{
\includegraphics*{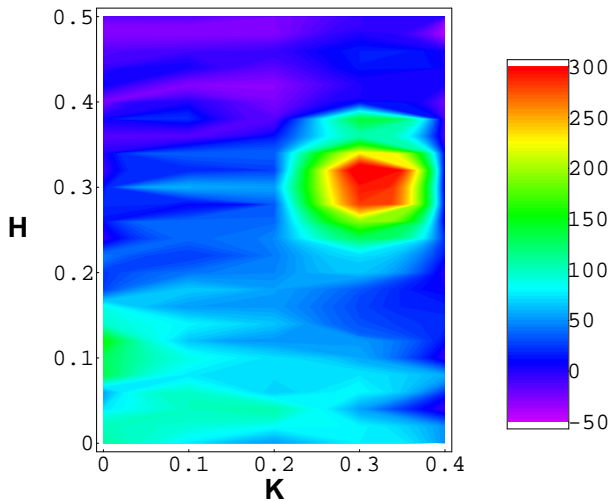}}
\caption{Mapping of the INS intensity by constant
energy scans with fixed $Q_y$ at 4.1meV energy transfer. 
The colour plot was obtained by adding symmetrical 
data after subtraction of the scattering angle dependent 
background.
}
\label{fig5}
\end{figure}

In order to address this problem we have mapped out the
intensity for ($Q_x$,$Q_y$,0) with $-0.5<Q_x<-0.0$  and
$0.6<Q_y <1.5$; these scans are shown in Fig. 5 after 
subtraction of the scattering angle dependent
background.
It is obvious that the incommensurate peak is by far the
strongest signal.
At low temperature one may roughly estimate
any additional signal to be at least a factor 6 smaller
than the nesting peak.
The analysis of such weak contributions
is quite delicate and demands a reliable subtraction of the
background.
Nevertheless the scans in the Fig. 5 indicate some
scattering closer to the zone center.
However, this contribution is not
sharply peaked at the $\Gamma$-point but forms
a broad square or a circle.
The signal roughly agrees with the prediction that
the \gbd ~ yields magnetic excitations near the
zone-center.

\begin{figure}[tp]
\resizebox{0.8\figwidth}{!}{
\includegraphics*{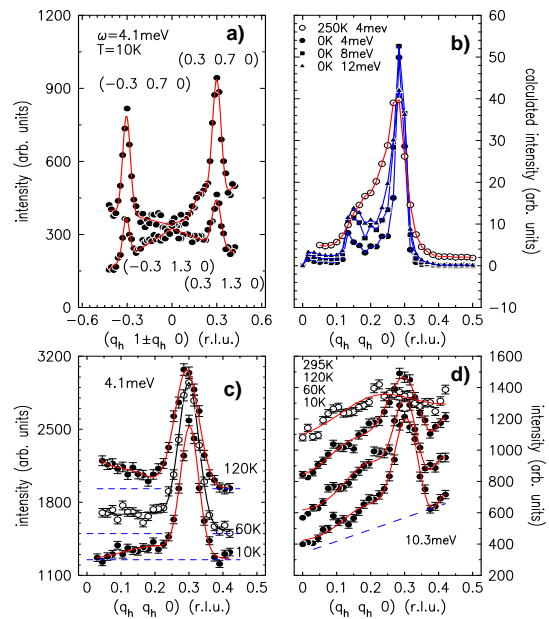}}
\caption{ a): Scans across the four incommensurate
spots surrounding ${\bf Q}$=(0,1,0) in diagonal 
direction. 
b) calculated intensity (corresponding to the 
imaginary part of the susceptibility 
multiplied with the thermal factor)
at different temperatures and energies.
c) sum of the four scans shown in part a) 
for different temperatures at \om =4.1meV.
c): Sum of the four scans shown in part a) 
at \om =10.3meV (at 295K not all the four scans 
have been performed).
.
}
\label{fig6}
\end{figure}

Since the sloping background is a major obstacle to analyze
the additional contributions we tried to compensate these
effects by scanning from ${\bf Q}$=(0 1 0) along the four diagonals,
which are illustrated in Fig. 1,
and by summing the four scans. The results are given in Fig. 6.
In the four single scans, Fig. 6a), 
one recognizes the nesting signal with
its intensity being determined by the form-factor and the
sloping background. The summed scans, see Fig. 6c), 
should have constant
background and the low temperature summed scan once more
documents that the nesting signal is by far the strongest one.
However, upon heating additional scattering contributions seem to
become enhanced in intensity in particular compared
to the nesting signal which decreases;
note that the Bose-factor will already
strengthen any signal by a factor three in the data in Fig. 6c).
Also, at higher energy the additional scattering seems to be
stronger as seen in Fig. 6d) (the background is strongly sloping
even in the sum due to a smaller scattering angle). 
The energy and temperature
dependence of the additional contribution corresponds
to that predicted by the full RPA analysis
for the \gbd ~ magnetic contribution 
shown in Fig. 6b). The 
spectral weight at ${\bf Q}$=(0.15,0.15,0) 
relative to that of the nesting feature
increases upon increasing temperature or energy
as it is expected for a signal directly related to the 
van Hove singularity.
Due to the agreement between the INS results and the RPA
calculations, we suggest to interpret the additional broad
scattering as being magnetic in origin; however, a polarized
neutron study would be highly desirable.
This scattering further might be relevant for a quantitative
explanation of the NMR-data \cite{imai,sidis}, in particular 
its temperature independent part.

\srruo ~ is not close to ferromagnetic order but substitution
of Sr through Ca yields such order for the concentration
\caruoef  \cite{13}. This doping effect was explained in a band structure
calculation \cite{38} as arising from a down shift in energy of the \gbd ~ 
pushing the van Hove-singularity closer to the Fermi-level.
In such samples the \gbd ~ magnetic scattering should therefore
become strongly enhanced. Indeed first INS studies
on these compounds reveal broad signals similar to the additional
scattering described above, but much stronger \cite{friedt}.
This strongly supports a magnetic interpretation of the scattering
in figures 5 and 6.

\subsection{Combined temperature and energy dependence of the
incommensurate signal } 

The energy and temperature dependence
has been studied in more detail by performing {\bf r}-scans
across the incommensurate position, since in this mode the
background is almost flat.
However, due to phonon contributions we could not extend the
measurements to energies higher than 12meV, which is slightly
below the lowest phonon frequency observed at \Qi 
~ \cite{35}. In particular
the large scans required to cover the broad magnetic
signal prevent any analysis within the phonon band frequency range on a
non-polarized thermal triple axis spectrometer.

\begin{figure}[tp]
\resizebox{0.6\figwidth}{!}{
\includegraphics*{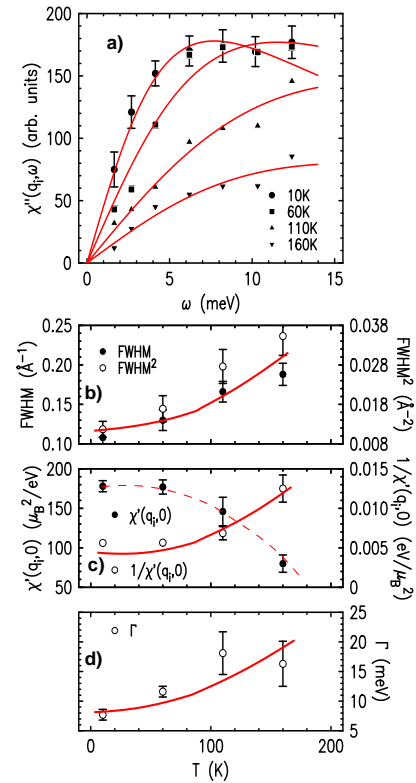}}
\caption{Observed imaginary part of the generalized susceptibility
as function of energy and temperature; lines are fits with a
single relaxor a). Temperature dependence of the averaged 
full width at half maximum (FWHM) corrected for the experimental 
resolution and temperature dependence
of the square of the FWHM  b). Temperature dependence
of the amplitude and its inverse c) and of the characteristic energy d) in the 
relaxor behavior fitted to part a). Lines in b-d) are guides to the eye.
}
\label{fig7}
\end{figure}

The results of the scans are given in Fig. 7.
At low temperature we find an energy spectrum in good agreement
to that published earlier \cite{sidis}.
In the range up to 12meV we observe at all temperatures
an energy independent peak width, which, however, increases
upon increase of temperature. 
For temperatures much higher than 160K the background 
considerably increases, see Fig. 2, and prevents a detailed
analysis within reasonable beam-time.
In Fig. 7b) we show the
temperature dependence of the peak width averaged over the
different energies which agrees well to the
results obtained from the single scans with less
statistics reported in reference \cite{sidis}. 
Even at the lowest temperature
the width of the signal remains finite.

The spectral functions have been fitted by a single 
relaxor behavior \cite{36}:
\begin{equation}
\chi "({\bf q_i}, \omega) = \chi '({\bf q_i}, 0)  \frac{\Gamma \omega}{\omega ^2 +
 \Gamma ^2}
\; \label{eq:3}
\end{equation}
where $\Gamma$ is the characteristic energy and
$\chi '({\bf q_i},0)$ the amplitude which
corresponds to the real part of the generalized susceptibility at
\om =0 according to the Kramers-Kronig relation.
The ${\bf Q}$-dependence of the signal may be described 
by a Lorentzian distribution with half width at half maximum
$\kappa$, but due to experimental broadening the constant energy scans
are equally well described by Gaussians.
At low temperature the spectrum
clearly exhibits a characteristic energy as seen in the
maximum of the energy dependence, but this maximum
is shifted to higher values upon heating.
Fig. 7c) and 7d) report the results of the least squares fits
with the single relaxor. Although the statistics is still
limited, the tendencies can be obtained unambiguously.
The height of the spectral functions,  $\chi '({\bf q_i},0)$,
rapidly decreases upon heating, which 
overcompensates the broadening of the signal in ${\bf Q}$-space.

Our finding that the characteristic energy in the range 6-9meV is well
defined only at low temperatures can be related with 
the far-infrared c-axis reflectance study by Hildebrand et al.
\cite{hildebrand}, since the optical spectrum 
shows a resonance in this energy range at low temperatures.
In reference \cite{sidis} we have compared the temperature dependence of the
incommensurate signal with that of the spin-lattice 
relaxation rate $T_1$ measured by both $^{17}$O and  $^{101}$Ru 
NMR experiments\cite{imai}.
These NMR-techniques probe the low energy spin fluctuations
($\omega \rightarrow 0$ with respect to INS measurements); furthermore,
they  integrate the fluctuations in ${\bf q}$-space.
$(1/T_1T)$ is related to the generalized susceptibility and the
INS results by \cite{nmr}:
\begin{equation} 
(1/T_1T) \simeq \frac{k_B\gamma_n^2}{(g\mu_B)^2} 
{  {\sum_{q}  { | A({\bf q}) |^2 \left.\frac{\chi "({\bf q},\omega)}
{ \omega} \right\vert_{\omega \rightarrow 0}}}}
\; \label{eq:1/T1}
\end{equation}
with $| A({\bf q}) |$ the hyper fine fields.
$(1/T_1T)$ corresponds hence to the slope of the spectral function
in Fig. 7a) times the extension of the signal in ${\bf Q}$-space.
The new data perfectly agrees to the former result, the
loss of the incommensurate signal upon heating may explain
almost entirely the temperature dependent contribution
to $(1/T_1T)$ \cite{imai}.

The temperature dependence of the magnetic excitation spectrum at the
incommensurate position may be analyzed within the results of the 
self consistent renormalization theory described in reference \cite{36}.
In an antiferromagnetic metal the transition is governed by a single 
parameter related to the Stoner-enhancement at the ordering wave-vector,
$\delta=1-I({\bf q_i})\chi^0({\bf q_i})$. The characteristic entities of the
magnetic excitations are then given by:

\begin{equation} 
\kappa ^2 \propto \delta 
\, \label{SCR1}
\end{equation}
\begin{equation} 
\Gamma \propto \delta 
\, \label{SCR2}
\end{equation}
\begin{equation} 
{1 \over \chi'({\bf q_i},0)} \propto \delta 
\ . \label{SCR3}
\end{equation}

When the system approaches the phase transition, the unique parameter
$\delta$ diminishes, which behavior should be observable in all three
parameters.
Equations (11-13)  imply a sharpening of the 
magnetic response in ${\bf q}$-space
as well as in energy and a divergence of the susceptibility at the ordering vector.
Fig. 7 qualitatively confirms this picture. All the relevant parameters, 
see the right scales in Fig.7, decrease towards low temperature.
Therefore, one may conclude that \srruo ~ is approaching the SDW transition
related to the nesting effects upon cooling.
However, all these parameters do not vanish completely but remain 
finite even at the lowest temperatures
in agreement with the well known fact that
\srruo ~ does not exhibit  magnetic ordering.
In particular the magnetic scattering remains rather broad
in ${\bf q}$-space implying a short correlation length of 
just 3-4 lattice spacings.
The temperature dependence of the magnetic excitations
corroborate our recent finding
that only a small amount of Ti is sufficient to induce 
SDW magnetic ordering \cite{srruti}.

Since \srruo ~ is close to a quantum critical point
it is tempting to analyze whether the excitation spectrum
is governed by some ${\omega \over T}$-scaling, as it has 
been claimed for the high temperature superconductor
\lsco ~\cite{aeppli,keimer,sachdev} and for CeCu$_{5.9}$Au$_{0.1}$ \cite{schroeder}.
One would expect that the susceptibility is given
by :
\begin{equation}
\chi"({\bf q_i},\omega,T)\propto T^{-\alpha}g({\omega\over T})\ .
\end{equation}

In Fig. 8 we plot the $\omega^{\alpha}\chi"({\bf q_i},\omega)$-data
of Fig. 7 and that obtained previously
as a function of temperature \cite{sidis} against
${\omega \over T}$ for $\alpha$=0.75 and 1.0. 
Only the data at higher temperatures agree with the 
scaling concept, 
demonstrating that \srruo ~ is not a quantum critical point.
The schematic
inset may illustrate the phase diagram, where the magnetic 
transition is determined by some parameter $r$ 
(external pressure or composition).
At the critical transition one would observe 
quantum criticality in the entire temperature range, 
whereas for r-values where the transition is suppressed
quantum criticality is observed only at higher temperatures.
One then may expect a cross-over temperature T$^*$ where 
the system transforms from an unconventional metal at 
high temperatures towards a Fermi-liquid at low temperatures
\cite{sachdev2}.
Only in the temperature range above T$^*$ the magnetic excitations
should exhibit the related ${\omega \over T}$-scaling.
Our data clearly shows that such scaling can be fitted to the 
data only for the three higher temperatures studied.
The description with the scaling concept seems to be
slightly better for the exponent $\alpha$=0.75.
The temperature dependent data suggests a crossover near 30K.
This cross-over agrees very well to that seen in
electronic transport properties, where well defined
Fermi-liquid behavior is only observed below about 25K \cite{maeno,2}.

\begin{figure}[tp]
\resizebox{0.9\figwidth}{!}{
\includegraphics*{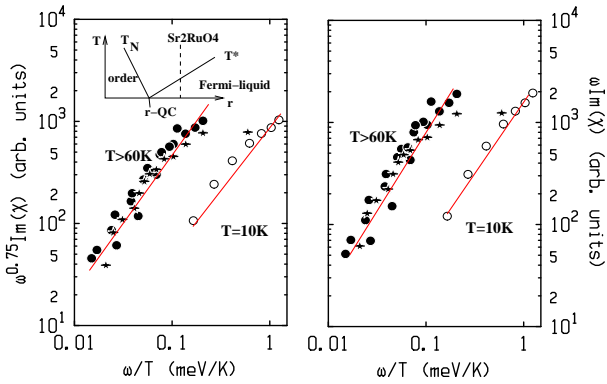}}
\caption{ $\chi"({\bf q_i},\omega,T)$
multiplied by $\omega^\alpha$  as a function 
of ${\omega \over T}$ in logarithmic axes;
the left and right parts correspond to $\alpha$=0.75
and 1 respectively. The inset gives a schematic representation
of a phase diagram close to a quantum critical point.
Solid and open circles correspond to the data shown in Fig. 7 
and the stars to that of our previous work \cite{sidis}.}
\label{fig8}
\end{figure}

\subsection{ Magnetic scattering in the superconducting phase }

As  emphasized by Joynt and Rice \cite{joynt}, the
wave-vector- and energy-dependent  spin susceptibility in superconductors
reflects directly the vector structure of the superconducting (SC) gap
function,
allowing
a complete identification of the SC order parameter symmetry. 
Inelastic neutron 
scattering experiments have the potential, in principle, 
to determine the superconducting order parameter.
In high-$T_c$ superconductors, the spin-singlet $d$-wave 
symmetry SC gap induces a striking modification 
of the spin susceptibility in the superconducting state. 
As a consequence, the so-called
"magnetic resonance peak" has been observed in the superconducting state of various 
copper oxide superconductors by INS\cite{rossat91,tl2201}.
Therefore, a similar experiment in \srruo would certainly be instructive about the SC gap 
symmetry.

The enhanced spin susceptibility has been calculated in \cite{27,39,39bis}, 
considering a spin-triplet p-wave superconductiviting state with {\bf d}({\bf k})={\bf z}($\rm k_x \pm i k_y$). Note that in such a case, the superconducting
gap is isotropic due to the particular shape of the 
Fermi-surface in \srruo .
For the wavevector \Qi, Kee et al. \cite{39} and Morr et al. \cite{27} predicted that below $T_c$ spectral weight is shifted 
from below twice the superconducting gap, $\Delta$, 
into a resonance like feature close to 2$\Delta$. 
Morr et al. find the resonance in the $zz$-channel yielding an
enhancement of the magnetic excitation intensity 
by a factor of 9 in the superconducting state as compared to the normal state\cite{27}.
The difference between the in-plane and out-of-plane susceptibility in the 
superconducting state arrises from the coherence-factor.

Similar theoretical framework is currently used to describe the spin excitation spectrum in spin-singlet 
HTSC cuprate superconductors\cite{OP}. 
Theoretical works show that the opening of
a d-wave order gap together with 
the exchange interaction lead to the appearence of a similar
resonant feature below 2$\Delta$ at the antiferronagnetic wavevector.
These theories successfully account for the observation
by INS of the magnetic resonance peak in the superconducting state.

\begin{figure}[tp]
\resizebox{0.9\figwidth}{!}{
\includegraphics*{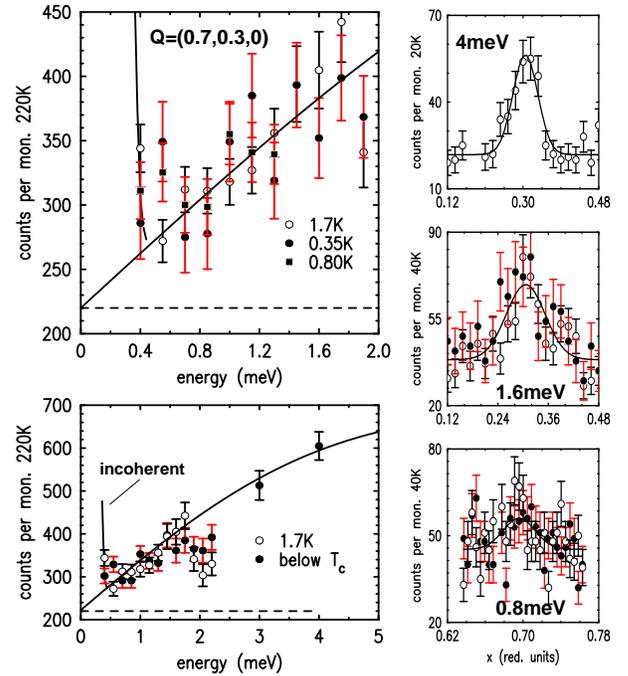}}
\caption{Results of the experiments performed across the
superconducting transition; 
filled and open symbols correspond to data taken
below and above $T_c$, respectively. The left part 
shows scans with constant ${\bf Q}$=(0.7,0.3,0) (in the lower
part the two low temperature scans were added), and the
right part shows constant energy scans.
The horizontal dashed line indicates the background.
}
\label{fig9}
\end{figure}

Using unpolarized INS,  one measures
the superposition of the out-of-plane and in-plane components of the susceptibility (see Eq.~\ref{eq:INSanisotropy}) and both components
are equally weighted, when performing the measurements $\rm Q_l$=0.
Thus, the predicted resonance feature should be observable,
at the value of  $\sim$2$\Delta$, if one obtains an 
experimental arrangement which allows to study the
inelastic magnetic signal in this energy range.
Due to the almost linear decrease of $\chi ''({\bf q_i},\omega)$
towards low energies, see Fig. 7, 
and due to the higher required resolution
which implies less neutron flux, 
these experiments are extremely time demanding.
We have analyzed the magnetic excitations in the 
superconducting phase on the cold triple axis spectrometer 
IN14 at the ILL using a two crystal assembly, the results
are shown in Fig.~\ref{fig9}.

The right part of Fig.~\ref{fig9}
shows a scan across the incommensurate peak at
\om= 4 meV, i.e. the range already studied with thermal 
neutrons. This signal can be determined with little 
beam time. Performing the same scan at 1.6 and 0.8 meV 
requires considerably more time but still exhibits
a well defined signal which seems not to experience 
any change in the superconducting state at T=0.35 K.
The results of constant-${\bf Q}$-scans at ${\bf Q}$=(0.7,0.3,0) 
~ are  shown in the left part of Fig.~\ref{fig9}. 
As there is no visible difference between the results obtained at
0.35 and 0.80 K (both in the SC state), we have added the two scans 
below $\rm T_c$ in the lower left part of Fig.~\ref{fig9}. 
Importantly and despite efforts to get rather high statistics, there
is no change visible in the energy spectra above and
below the superconducting transition in the energy 
range of the superconducting gap.
The spin susceptibility is not modified appreciably across the 
superconducting transition; 
our data even do not show any opening of a gap. 
One can describe the energy dependence presented in Fig. 9
with the single relaxor using the same 
fitted parameters as the low temperature data in Fig.~\ref{fig5}.a.
However, the detailled shape of the spin suceptibility (Fig.~\ref{fig9}) 
does not exactly 
match such a simple linear behavior but rather seems to indicate some anomaly
near 2meV
which requires further experimental work.

There is actually little known about the value of the superconducting gap
in \srruo ~. Laube et al. \cite{40} have reported an Andreev reflection study
where the opening of the gap is clearly visible in an astonishingly
large energy range. The quantitative analysis of the spectra is 
quite involved; assuming a p-wave order parameter Laube et al.
obtain 2$\Delta$=2.2 meV which may be compared to the
value expected within BCS-theory 
$2\Delta$= 3.55 k$_B$$T_c$ = 4.97 K = 0.43 meV.
Our data shows that there is no change in the excitation spectrum
for energies well below the reported value of 
2$\Delta$, but it has not been possible to investigate
the lowest energies due to the strong elastic incoherent 
signal. With further increased resolution ($k_f=1.3$\AA$^{-1}$)
we have scanned the energy range 0.3--0.7 meV at 0.80 K again without 
evidence for a resonant feature.
Furthermore, the comparison of two scans at constant ${\bf Q}$=(1,0,0)
did not yield any difference below and above $\rm T_c$ meaning 
no ferromagnetic spin susceptibility enhancement.

The theory presented by Morr et al. \cite{27} should be considered
as being quite reliable in the case of \srruo, since the RPA
approach to the magnetic excitations is so successful in the
normal state, and since \srruo ~ exhibits well defined 
Fermi-liquid properties at temperatures below 25\ K.
Therefore, the data in Fig.~\ref{fig9} gives strong evidence against
a simple p-wave order parameter in \srruo ~ with a maximum
value of the gap of the order of the reported value \cite{40}.
However, in the meanwhile there are several indications
that the order parameter is more complex.
The recent specific heat data on the highest quality
single crystals \cite{21} points to the existence of line nodes 
in the gap function which were then shown to be aligned
parallel to the a,b-plane (horizontal line nodes).
Such line nodes were explained  by Zhitomirsky and Rice
through a proximity effect between the active \gbd ~
and the more passive one-dimensional bands \cite{24}.
A modulation of the gap function along the c-direction
will wipe out the resonance predicted for the non-modulated
p-wave gap, since for the ${\bf Q}$-position analyzed, (0.7 0.3 0),
the electron hole excitation involves parts of the 
Fermi surface which are fully, partially or not gapped at all.
In this sense the absence of any temperature dependence
in the magnetic excitation spectrum is consistent
with the presently most accepted shape of the 
gap function. Further theoretical as well as experimental
studies are required to clarify the possibility of a resonant
feature at other positions in $({\bf Q},\omega)$-space.

\section{Conclusions}

Using assemblies of several crystals of \srruo ~ we have analyzed the
magnetic excitations by INS.
The incommensurate signal arising from the nesting between
the one-dimensional bands shows an asymmetry which is well 
explained by the full RPA analysis as a contribution 
mainly from the \gbd . 

The energy dependence of the
incommensurate signal varies with temperature and exhibits
a general softening of the spectrum upon cooling.
This behavior indicates that \srruo ~ is approaching the 
corresponding SDW instability at low temperature even though 
this compound is not at a quantum critical composition.
This interpretation is confirmed by the fact that 
the generalized susceptibility exhibits some $\omega / T$-scaling only
above $\sim$30\ K, i.e. in the temperature range 
where also the transport properties indicate 
non-Fermi-liquid behavior.

The analysis of magnetic excitations besides the nesting
ones shows only minor contributions. There is some 
evidence for additional magnetic scattering closer to the 
zone-center but still not peaking at the zone-center.
This interpretation gets support from the fact that similar scattering is 
observed in nearly ferromagnetic \caruoef ~ and from 
various RPA calculations which find excitations mainly related 
to the \gbd ~ in this ${\bf q}$-range.

The magnetic excitations in \srruo ~ may be compared 
to the distinct types of magnetic order which have been
induced by substitution. The dominant excitations reflect
the SDW reported to occur in \srrutio ~ at small Ti
concentrations. The less strong excitations situated
more closely to the zone-center and most likely related 
to the \gbd ~ become enhanced through Ca-substitution
which drives the system towards ferromagnetism, but only
for rather high Ca-concentration.  

Upon cooling through the superconducting transition we do
not observe any change in the magnetic excitation
spectra, which - combined with recent calculations \cite{27}-
indicates that the order-parameter in \srruo ~ does not possess
simple p-wave symmetry. 
These experimental findings are still in agreement with
a p-wave order parameter modulated by horizontal line nodes.

{\bf Acknowledgements.} We would like to thank O. Friedt, H.Y. Kee,
D. Morr and R. Werner
for stimulating discussions and P. Boutrouille (LLB) and S. Pujol (ILL)  
for technical assistance. Work at Cologne University was supported 
by the Deutsche Forschungsgemeinschaft through the 
Sonderforschungsbereich 608. Work at at Kyoto was supported by 
a grant from CREST, Japan Science and Technology Corporation.

* electronic mail : braden@ph2.uni-koeln.de


\begin{references}
\bibitem{formfac} {\it International Tables of Crystallography}, Vol. C, 
eds A. J. C. Wilson, Kluwer Academic publisher (1995);
note, that the form factor for Ru$^{4+}$ is not tabulated.
\end{references}
\end{document}